\documentclass[twocolumn]{article}
\usepackage{epsfig,amssymb,amsmath,setspace,cmbright, hyperref}

\usepackage{amssymb}

\begin{document}

\title{\begin{flushleft}{\bf Can the Internet cope with stress?\hfill\\[-.2 cm]}\end{flushleft}}
\date{\begin{flushleft}{Andreas Martin Lisewski\footnote{Email:{\tt lisewski@bcm.edu}}\\
{\scriptsize Baylor College of Medicine, One Baylor Plaza, Houston, TX 77030, USA}\\{\small April 30, 2007}}\end{flushleft}}
\author{}
\maketitle


\begin{abstract}
{\begin{flushleft}\small 
When will the Internet {\it wake up} and become aware of itself? In this note the problem is approached by asking an alternative question: Can the Internet cope with stress? By extrapolating the psychological difference between coping and defense mechanisms a distributed software experiment is outlined which could reject the hypothesis that the Internet is not a conscious entity.
\end{flushleft}} 

\end{abstract}

{\it\small ``I know that you and Frank were planning to disconnect me, and I'm afraid that's something I cannot allow to happen.''}
{\begin{flushright}
{\small --- HAL 9000}\\[1cm]
\end{flushright}} 

\section*{Introduction and background}

The idea of a self-awareness and consciousness developing global computer network has migrated from science fiction to the attention of neuoscientists. Terrence Sejnowski has recently readdressed this thought \cite{sejnowski} claiming that the Internet's rapid growth, its communication design and architecture along with some of its functional measures, such as storage capacity and bandwidth, resemble neurobiological aspects or are not far from values representative of the human brain. Although still speculative, it seems possible that the idea can be eventually formulated as a valid scientific hypothesis, which, however, cannot be decided from today's scientific knowledge in neuroscience, according to Sejnowski. This inability likely stems from the fact that no scientifically accepted and objective procedure has been found that would allow a detection of consciousness in any given object or organism, hence from the current lack of a convincing consciousness test \cite{aleksander2001, harnad2001}.
\medskip

The purpose of this note is to add to the problem a psychological perspective, which could lead to a testable strategy regarding the corresponding null hypothesis, i.e., that the current Internet has no detectable form of consciousness. The main argument is that the null hypothesis can be rejected on the ground of two distinct stress and problem situation adaptation processes known to psychology: coping and defense. This argument is based on criteria that differentiate between defense and coping and include the conscious/unconscious status and the intentional/non-intentional nature of the process. Phebe Cramer, in her review on coping and defense \cite{cramer}, summarizes that ``coping mechanisms involve a conscious, purposeful effort, while defense mechanisms are processes that occur without conscious effort and without conscious awareness (i.e., they are unconscious). Also, coping strategies are carried out with the intent of managing or solving the problem situation, while defense mechanism occur without conscious intentionality; the latter function to change internal psychological state but may have no effect on external reality,[...]''.
\medskip

Even though coping and defense were introduced as psychological dimensions, an extended interpretation is here proposed where both aspects are brought to the context of sufficiently complex communication networks, such as the current Internet, which may have a potential to develop consciousness. As a system, the Internet can be characterized with both an internal state and an external reality. The internal state consists of computer programs and communication protocols that regulate and control the network, and of more abstract entities such as the network's connectivity, scalability and redundancy. On the other hand, a relevant part of the external reality are human agents who physically interact over human-computer interfaces with the Internet. Although an interpretation of coping and defense outside of psychology bears a risk of categorical error, it is notable that this global computer network has rapidly emerged into a complex system capable of defense mechanisms in the presence of external stress. Its dynamic, decentralized, distributed and redundant internal structure has made it adaptable and resilient to adverse situations triggered by transient external events such as operator errors, power outages, natural disasters, and forceful attacks on infrastructure \cite{istress}. It can be therefore hypothesized that, given an external stressor, coping with stress would also become a possibility, leading to an intentional change of external reality by addressing the cause of stress.
\medskip

To arrive at an empirically testable procedure, it is conjectured that stress can be inflicted on the Internet by a concerted synchronous shutdown of a sufficiently large number of connected computers (hosts) under the designated control of human agents (users). Actively disconnecting computers from the Internet means reducing its storage capacity, lowering redundancy and connectivity, and diminishing the level of external interaction by temporarily discarding human-computer interfaces, thus causing systemic stress. Such concerted action would require a foregoing planning stage followed by a directed shutdown event both communicated and executed by a group of volunteers. In a hypothetical response, a coping Internet would act to prevent the shutdown by trying to interfere with external reality of the users, hence to change the course of events and to effectively reduce stress.\footnote{It can only be speculated about the Internet's specific coping strategies. For example, it is imaginable that it subtly draws the user's attention away from her/his original plan through transient audio-visual stimuli.} Coping would result in an intentional conflict between a group of users, who plan to execute a large-scale shutdown and to impose stress, and the Internet itself, which intends to prevent this action. The decisive question is whether a coping Internet could intentionally dissuade users from their aim by interfering with their reality. A defending Internet, on the other hand, would affect only its internal state, for example through adjustments of its communication protocols or through changes in connectivity and redundancy, and no conflict situation would arise. Here, from a user perspective, no hindrances occur and the plan can be implemented straight forward.
\medskip

In what follows, a simple interaction model between human users and the Internet is introduced, representing coping and defense in the presence of external stress. To empirically test the model, a specific experimental protocol is outlined.

\section*{Experiment outline}

Human agents $\mathbb{H}$ can interact as users with the Internet $\mathbb{I}$ by sending information toward and, in response, by retrieving information from it. Three types of interaction are distinguished: if a directed action $\mathbb{H}\rightarrow\mathbb{I}$ initiated by $\mathbb{H}$ causes a stressful or an adverse situation for $\mathbb{I}$ (here, an electrical shutdown $S$ of many participating constituents of $\mathbb{I}$), then a defending Internet internalizes and adapts to this action ($\mathbb{I} \rightarrow \mathbb{I}$), while a coping internet additionally reacts and influences external reality to inhibit the action of the stressor ($\mathbb{I}\dashv\mathbb{H}$). These alternatives are denoted as $D$ and $C$, respectively, so that a realization of $C$ would be an indicator of coping and, in turn, of a conscious act.
\medskip

A sufficiently large and synchronized shutdown could be realized with the help of a computer program $\mathbb{R}$, the so-called client, which is distributed among many users.  The current size of the Internet requires a relatively large number $N$ of users who operate $\mathbb{R}$ on hosts which are physically accessible to them. By February 2007, the number of hosts estimated by the Internet Systems Consortium (ISC) domain survey reached 500 million \cite{isc}---a reasonable lower bound, because one single physical computer can carry multiple (virtual) hosts and many hosts exist which could not be reached over the network at the time of the survey. Thus even a large $N$, say five million, would affect only $1 \%$ of the current Internet. This level of user particaption seems nevertheless realistic since it compares to other popular distributed computing projects, such as {\it SETI@home}, where the number of program copies for active project members has grown over five million in the year 2005 \cite{seti}.
\medskip

During experiment, the client $\mathbb{R}$ has to accomplish two tasks. Firstly, in the so-called calibration phase, it periodically reports the number of participating users. For that it applies a time synchronization  protocol---such as the Network Time Protocol \cite{ntp}--to simultaneously perform a series of tests at successive times $\{t_1, t_2,\ldots,t_n\}$ and to report the test outcome to a so-called counter $\mathbb{Z}$, a central host set up as an analyzer. Beginning at each time $t_i$, the client starts the test by asking its user whether he or she would refrain from any mechanical interaction with the host during the following $\Delta\tau=15$ minutes. When agreed to participate in the test, the user is instructed not to use any of the host's mechanical human-computer input devices, such as alphanumeric keyboard or {\it mouse}, during that time. The client controls if this instruction has been followed upon agreement, and it sends a unique message to $\mathbb Z$ to report the control outcome. The period between messages $\Delta t = (t_i+\Delta\tau) - (t_{i-1}+\Delta\tau)$ can be set conveniently, e.g., to 24 hours or seven days, and the message itself can be a random string of letters which is unique at every time but the same for every client. The counter $\mathbb{Z}$ registers the total number of incoming messages for every $ t_i+\Delta\tau$, thus estimates the number $N_i$ of participating users who stopped interaction with their hosts, according to the control. This registration process yields a discrete distribution of participation numbers $\{N_1, \ldots, N_n\}$, and given that values $N_i$ do not strongly fluctuate, e.g., all $N_i$ remain within an order of magnitude, they define a sample mean $\bar N$ and standard deviation $\sigma$. It is also assumed that the normalized distribution $\{N_1, \ldots, N_n\}$  can be reasonably approximated by a normal distribution. 
\medskip

Secondly, in the execution phase, all users $\mathbb{H}$ attempt at time $t_*=t_n+\Delta t$ the shutdown procedure $S$ on their physically accessible hosts, hence disconnecting them from the Internet for the time of $\Delta\tau = 15$ minutes. After reconnection with the Internet a control is made and every client $\mathbb{R}$ sends a unique message to $\mathbb{Z}$ to certify that the corresponding host has been switched off and disconnected as planned.\footnote{This certification can follow from an analysis of the host's system log file, for example.} As a consequence, the number of messages $N_*$ received by $\mathbb{Z}$ represents a raw number of disconnected hosts which can be statistically standardized by a $z$-score, $z=(N_*-\bar N)/\sigma$. A crucial but reasonable assumption is that the user's effort to shutdown and disconnect the host in the execution phase is comparable with the effort of not using the computer over the period $\Delta\tau$ during the calibration phase of the experiment. Therefore, in the absence of any coping mechanism, $N_*$ would be comparable to $\bar N$ and $z$ would be close to zero. According to the experiment hypothesis, a positive or absolutely small  $z$-score indicates a defending Internet that had no direct influence on the specific cause of stress (negative outcome). On the other hand, if $\mathbb{I}$ has the capacity to hinder $\mathbb{H}$ from executing this process, that is, if $\mathbb{I}$ can realize the alternative $C$ by coping with stress,  then it is expected that $N_*$ would be less than $\bar N$, resulting in a negative $z$-score (positive outcome). A negative $z$-score would therefore reject the corresponding null hypothesis $D$ with a probability $1-P(z)$, where $P(z)$ is the cumulative percentile of the normal distribution corresponding to $z$. 
\medskip

In case of a positive outcome, it would also be helpful to examine the specific circumstances which prevented users from switching off and disconnecting computers. Here the client program could additionally offer a questionnaire survey to analyze possible reasons that lead to a deviation from the original plan. This could shed some light on the Internet's disposition to intentionally and specifically interfere with user reality. 
\medskip

From this simple experimental protocol it appears that it may not be impossible to provoke a coping response from the Internet and to measure its effects. 
\bigskip

The author thanks K. Koepsell, M. Meissner and T. von Merveldt for valuable comments.

\end{document}